\documentclass[conference,letterpaper]{IEEEtran}

\usepackage[utf8]{inputenc} 
\usepackage[T1]{fontenc}
\usepackage{url}
\usepackage{ifthen}
\usepackage{cite}
\usepackage[cmex10]{amsmath} 
\usepackage[utf8]{inputenc} 
\usepackage[T1]{fontenc}
\usepackage{url}
\usepackage{ifthen}
\usepackage{cite}
\usepackage[cmex10]{amsmath} 
\usepackage{epsfig,epsf,rotating,setspace,latexsym,amssymb,amsfonts,bm,theorem,epstopdf,authblk,bbm,mathrsfs,graphicx,caption, tabularx,dsfont}
\usepackage{subcaption}
\usepackage{algorithm}
\usepackage[noend]{algpseudocode}
\usepackage{color}
\usepackage{mathtools}
\usepackage{multirow}
\usepackage{soul}

\algrenewcommand\algorithmicforall{\textbf{foreach}}
\algrenewcommand\algorithmicindent{.8em}

\newtheorem{theorem}{Theorem}
\newtheorem{lemma}{Lemma}
\newtheorem{corollary}{Corollary}
\newtheorem{remark}{Remark}
\newtheorem{definition}{Definition}

\newtheorem{example}{Example}
\newenvironment{Proof}[1]{\medskip\par\noindent{\bf Proof:\,}\,#1}{{\mbox{\,$\blacksquare$}\par}}

\title{Local Private Information Retrieval: A New Privacy Perspective for Graph-Based Replicated Systems}

\author{Shreya Meel\qquad  Mohamed Nomeir \qquad Sennur Ulukus\\
\normalsize Department of Electrical and Computer Engineering\\
\normalsize University of Maryland, College Park, MD 20742 \\
\normalsize \it  smeel@umd.edu  \qquad mnomeir@umd.edu \qquad ulukus@umd.edu}
 
\begin{document}

\maketitle

\begin{abstract}
    We rethink the definition of privacy in multi-server, graph-replicated private information retrieval (PIR) systems, and introduce a novel setting where the user's privacy is governed by the servers' storage structure. In particular, while retrieving a message from a server, the user is concerned with hiding their desired message index from the server, only if the server stores the corresponding message. We coin this privacy requirement as \emph{local user privacy} and the resulting PIR problem as \emph{local PIR} on the graph. Our goal is to measure the gain in communication efficiency of local PIR, compared to that of canonical PIR, by establishing its capacity, i.e., the maximum number of message symbols retrieved, per downloaded symbol. To this end, we observe a remarkable gain in the local PIR capacity of graphs, that are disjoint union of distinct graphs, which is multiplicative, compared to the PIR capacity, when the individual graphs are identical. For connected graphs, we propose schemes to establish capacity lower bounds for edge-transitive and bipartite graphs, which are greater than the best-known PIR capacity bounds. Finally, we derive the exact local PIR capacity for the cyclic graph, and the path graph with an odd number of vertices.
\end{abstract}

\section{Introduction}
Private information retrieval (PIR) \cite{chor} is the framework where a user downloads their desired message from an indexed database, while hiding its index from the database. This problem is motivated by crucial applications where knowing the user's request reveals the user's interest, which is detrimental to the user's privacy. The PIR system model consists of $N$ servers storing $K$ messages, where the user communicates with the servers to retrieve one of these messages privately. PIR has garnered renewed interest among the information-theoretic community, with the goal of studying its capacity, under various settings. The capacity of PIR, defined as the maximum ratio between the message size, and the amount of downloaded information, was derived in \cite{SJ17} for the fully-replicated database setting, where the $N$ servers store all $K$ messages. This was extended to address other security concerns, like, database security \cite{c_spir,  skoglund_mds_spir, zhusheng_spir_pir}, server collusion \cite{colluding, pir_spir_adversaries, coded_colluding_2017, arbitrarycollusion,  csa, nomeirasymmetric,semantic_tpir}, eavesdroppers \cite{sun_eaves, nan_eaves, C_SETPIR}, malicious servers \cite{byzantine_tpir,  nomeir_asymp_bspir}. 

However, as the amount of data keeps growing at a rapid pace, the assumption that all messages are stored at all servers is becoming expensive and unrealistic. Moreover, geographical and access constraints may restrict the storage of some data at certain servers. To address this, PIR with non-fully-replicated databases was introduced in \cite{graphbased_pir,  BU19}, where the storage of messages at the servers is modeled by a hypergraph. Specifically, the servers are represented by vertices, and each message is represented by a hyper-edge connecting to vertices, corresponding to the servers where the message is replicated. This model was specialized to simple graphs, where each edge consists of exactly two vertices, i.e., each message is stored in two servers \cite{SGT23,  YJ23,  our_journal2025,  krishnan_graph, gePIR},  and to $r$-multigraphs \cite{meel_multi_pir,  gePIR} (where each edge is replaced by $r$ parallel edges), and capacity bounds were established for well-known graphs and multigraphs. 

Although hypergraph-replicated PIR accommodates partial replication of messages, the privacy requirement has been kept the same as the one in the full-replication formulation in \cite{SJ17}. Consequently, it mandates the retrieval of all messages to be uniformly private at all servers, against all message indices, irrespective of whether the given message is stored at the server or not. However, this may be too restrictive towards the goal of preventing user-profiling. From a practical standpoint, if a server does not store a message and is oblivious to its contents, knowing that the user has requested that message is unlikely to breach the user's privacy. From this viewpoint, we introduce a new perspective to graph-replicated PIR, by studying \emph{local PIR}, wherein the only message indices that are kept private from a given server are those of the messages that are stored in it. This relaxes the privacy constraint, and increases the capacity for a given graph. Note that, our formulation extends seamlessly to the fully-replicated setting, where local PIR is equivalent to the original PIR formulation.

In this work, we focus on simple graphs, and derive \emph{local PIR capacity} bounds for well-known graph families. First, if a graph is a disjoint union of multiple graphs, we establish a direct connection between the capacity of the overall graph, and that of each individual graph. For connected graphs, i.e., graphs in which there is a path between every pair of vertices, we propose achievable local PIR schemes for edge-transitive and bipartite graphs. These two graph families subsume the graphs whose PIR capacity is known exactly, or up to a constant factor, enabling us to show that local PIR capacity is greater. Moreover, we derive the local PIR capacity for the cyclic graph with $N$ vertices to be $\frac{1}{2}$, i.e., independent of the number of servers, while its PIR capacity is $\frac{2}{N+1}$\cite{BU19}. For the path graph, we establish the local PIR capacity exactly as $\frac{N-1}{2N-4}$ for odd $N$, compared to $\frac{2}{N}$ as its PIR capacity\cite{our_journal2025}.

\section{Problem Formulation}
In graph-replicated PIR systems, each server stores only a subset of the message set $\mathcal{W} =\{W_1,\ldots,W_K\}$. Each message $W_k\in\mathcal{W}$ is replicated exactly twice and stored on two distinct servers in $[N]=\{1,2,\ldots,N\}$. We represent the database system by a simple, undirected graph $G=(V,E)$, where $V=[N]$, and $|E|=K$. Hence, each vertex represents a server and each edge $\{i,j\}\in E$ incident with servers $i$ and $j$, is assigned the index of the message replicated at those servers. Each message $W_k$ is an independent vector of $L$ independent, and uniformly random symbols from a finite field $\mathbb{F}_q$, with 
\begin{align}
    H(W_1,\ldots,W_K) = \sum_{k=1}^K H(W_k) = K L
\end{align}
in $q$-ary units. We denote the subset of messages at server $n$ as $\mathcal{W}_n$. Similarly, we denote the \emph{message indices} in $\mathcal{W}_n$ as 
\begin{align}
    \mathcal{I}_n=\{\ell:W_\ell\in \mathcal{W}_n\}\subseteq [K].
\end{align}

Let $\theta\in [K]$ be the message index required by the user, where $\theta$ is uniformly random over $[K]$. In PIR, to retrieve the message $W_{\theta}$, the user sends the query $Q_n^{[\theta]}$ to server $n\in [N]$. The set of queries for all servers across all $\theta$, denoted by $\mathcal{Q} = \{Q_n^{[\theta]}, n\in [N], \theta\in [K]\}$ is independent of the stored messages at the servers, i.e.,
\begin{align}\label{query-msg_indep}
    I(\mathcal{Q};W_1,\ldots,W_K) = 0.
\end{align}
Recall that, in the canonical formulation of graph-replicated PIR, the query sent to server $n$ must satisfy
\begin{align}
    I(\theta;Q_n^{[\theta]}, W_1,\ldots,W_K) = 0, \quad n\in [N].
\end{align}
In contrast, we introduce the \emph{local user privacy} requirement where the user's privacy against server $n$ concerns only with the messages stored in $\mathcal{W}_n$. That is, when the user requests for one of the messages in $\mathcal{W}_n$, server $n$ learns nothing beyond knowing that $\theta\in \mathcal{I}_n$. For local user privacy, $Q_n^{[\theta]}$ must satisfy
\begin{align}\label{eq:user privacy new}
    I(\theta;Q_n^{[\theta]},\mathcal{W}_n| ~\theta \in \mathcal{I}_n)&=0, \quad   n\in [N]. 
\end{align}
It is important to note that, in the canonical graph-replicated PIR system also requires $I(\theta;Q_n^{[\theta]},\mathcal{W}_n|~\theta\notin \mathcal{I}_n)=0$, in addition to \eqref{eq:user privacy new}, which is the key difference from our local privacy definition.

Upon receiving the query, server $n$ responds with the answer $A_n^{[\theta]}$, which is a deterministic function of $Q_n^{[\theta]}$ and $\mathcal{W}_n$, i.e.,
\begin{align}\label{answer_function}
    H(A_n^{[\theta]}|Q_n^{[\theta]},\mathcal{W}_n) = 0.
\end{align}
As in the usual PIR, from all the received answers, i.e., $A^{[\theta]}_{[N]}=\{A_1^{[\theta]},\ldots,A_{N}^{[\theta]}\}$, and for any realization of $\theta \in [K]$, the message $W_{\theta}$ should be exactly recovered by the user, i.e.,
\begin{align}\label{eq:decodability}
    H(W_{\theta}|A_{[N]}^{[\theta]}, \mathcal{Q}) = 0.
\end{align}
A \emph{local PIR} scheme $\Pi$ on a graph $G$ is identified by the set of queries $\mathcal{Q}$, and the corresponding set of answers $\mathcal{A} = \{A_n^{[\theta]}, n\in [N], \theta \in [K]\}$ which satisfy \eqref{query-msg_indep}, and \eqref{eq:user privacy new}-\eqref{eq:decodability}. For $\theta=k$, let the total number of $q$-ary symbols downloaded be $D_k = \sum_{n=1}^N H(A_n^{[k]})$.
Further, let $D$ represent the download cost random variable, associated with the random index $\theta$. Then, given the graph $G$, the rate of a local PIR scheme $\Pi$ is defined as,
\begin{align}\label{eq:rate_expected}
    R^{\Pi}(G) = \frac{L}{\mathbb{E}[D]}= \frac{KL}{\sum_{k=1}^K D_k}.
\end{align}
The capacity of local PIR for $G$ is the supremum over all achievable rates, i.e., $C(G)=\sup_{\Pi} R^{\Pi} (G)$.

\begin{remark}
    Due to the relaxed privacy constraint in \eqref{eq:user privacy new}, we have $C(G)\geq C_{PIR}(G)$, where $C_{PIR}$ is the PIR capacity with the standard (non-local) privacy constraint.  
\end{remark}

\section{Motivating Examples}
In this section, we demonstrate the main idea of our formulation through local PIR schemes on well-known graphs.

\begin{example}[Cyclic graph]
Consider the cyclic graph $\mathbf{C}_N$ for $N=4$ with the storage $\mathcal{W}_1 = \{W_1,W_4\}$, $\mathcal{W}_2 = \{W_1,W_2\}$, $\mathcal{W}_3 = \{W_2,W_3\}$ and $\mathcal{W}_4 = \{W_3,W_4\}$.
Assume $L=2$ and let $W_1=(a_1,a_2)$, $W_2=(b_1,b_2)$, $W_3 = (c_1,c_2)$ and $W_4=(d_1,d_2)$ be the messages, after the user independently permutes the message symbols, uniformly at random. Table~\ref{tab:C4 answers} illustrates the local PIR answers. Note that, privacy against server $n$ is maintained for indices in $\mathcal{I}_n$. The rate of the scheme is $\frac{1}{2}$, which is actually the capacity of this setting, while $C_{PIR}(\mathbf{C}_4)=\frac{2}{5}<\frac{1}{2}=C(\mathbf{C}_4)$. 

\begin{table}[htbp]
    \centering
    \begin{tabular}{|c|c|c|c|c|}
    \hline
        & server 1& server 2& server 3& server 4\\
        \hline
        $\theta = 1$& $a_1+d_1$ & $a_2+b_1$ & $b_1$ & $d_1$\\
        \hline
        $\theta = 2$& $a_1$ & $a_1+b_1$ & $b_2+c_1$ & $c_1$\\
        \hline
        $\theta = 3$& $d_1$ & $b_1$ & $b_1+c_1$ & $c_2+d_1$\\
        \hline
        $\theta = 4$& $a_1+d_1$ & $a_1$ & $c_1$ & $c_1+d_2$\\
        \hline
    \end{tabular}
    \caption{Retrieval scheme for $\mathbf{C}_4$.}
    \label{tab:C4 answers}
\end{table}
\end{example}

\begin{example}[Complete graph]
In a complete graph $\mathbf{K}_N$, an edge exists between each vertex pair, thus the total number of edges is $\binom{N}{2}$. Equivalently, a message is stored at every pair of servers. Consider the case of $N=4$, i.e., $\mathbf{K}_4$, where the storage is $\mathcal{W}_1 = \{W_1,W_2,W_3\}$, $\mathcal{W}_2= \{W_1,W_4,W_5\}$, $\mathcal{W}_3 = \{W_2,W_4,W_6\}$ and $\mathcal{W}_4= \{W_3, W_5, W_6\}$. Assume $L = 4$, and let the permuted message symbols be $W_1=(a_1,\ldots,a_4)$, $W_2=(b_1,\ldots,b_4)$, $W_3=(c_1,\ldots,c_4)$, $W_4 = (d_1,\ldots,d_4)$, $W_5 = (e_1,\ldots,e_4)$ and $W_6 = (f_1,\ldots, f_4)$. The answers of a local PIR scheme are shown in Table~\ref{tab:K4 answers}. The rate of the scheme is $\frac{2}{5}$, which is greater than the upper bound on the PIR capacity, which is $0.35\leq C_{PIR}(\mathbf{K}_4)\leq 0.3529$ by \cite{gePIR}.

\begin{table}[htbp]
    \centering
    \begin{tabular}{|c|c|c|c|c|}
    \hline
    & server 1 & server 2 & server 3& server 4\\
    \hline
    \multirow{3}{*}{$\theta =1$} & $a_1+b_1$ & $a_3+d_1$ & $b_1$ & $c_1$\\
    & $a_2+c_1$  &  $a_4+e_1$& $d_1$ & $e_1$\\
    & $b_2 +c_2$ & $d_2+e_2$ & &\\
    \hline
    \multirow{3}{*}{$\theta =2$} & $a_1+b_1$ & $a_1$ & $b_3+d_1$ & $c_1$ \\
    & $a_2+c_1$ & $d_1$ & $b_4+f_1$ & $f_1$\\
    & $b_2+c_2$ & & $d_2+f_2$ &\\
    \hline
     \multirow{3}{*}{$\theta =3$} & $a_1+b_1$ & $a_2$ & $b_2$ & $c_3+e_1$ \\
    & $a_2+c_1$ & $e_1$ & $f_1$ & $c_4+f_1$\\
    & $b_2+c_2$ & & & $e_2+f_2$\\
    \hline
    \multirow{3}{*}{$\theta =4$} & $a_1$ & $a_1+d_1$ & $b_1+d_3$ & $e_2$ \\
    & $b_1$ & $a_2+e_1$ & $b_2+f_1$ & $f_2$\\
    &  & $d_2+e_2$ & $d_4+f_2$ &\\
    \hline
     \multirow{3}{*}{$\theta =5$} & $a_2$ & $a_1+d_1$ & $d_2$ & $c_1+e_3$ \\
    & $c_1$ & $a_2+e_1$ & $f_2$ & $c_2+f_1$\\
    &  & $d_2+e_2$& & $e_4+f_2$\\
    \hline
    \multirow{3}{*}{$\theta =6$} & $b_2$ & $d_2$ & $b_1+d_1$ & $c_1+e_1$ \\
    & $c_2$ & $e_2$ & $b_2+f_1$ & $c_2+f_3$\\
    &  & & $d_2+f_2$& $e_2+f_4$\\
    \hline
    \end{tabular}
    \caption{Retrieval scheme for $\mathbf{K}_4$.}
    \label{tab:K4 answers}
\end{table} 
\end{example}

\begin{example}[Star graph]
A star graph $\mathbf{S}_N$ consists of a central server (say server $N$) storing all the messages, i.e., $\mathcal{W}_N = \{W_1,\ldots,W_{N-1}\}$, and $N-1$ dedicated servers, each storing a single message, i.e., $\mathcal{W}_n = \{W_n\}$ for $n\in [N-1]$. Then, we have the following local PIR scheme. For $\theta \in [N-1]$, we query only server $\theta$ for $W_{\theta}$, and do not download anything from server $N$. For each $\theta=k$, $D_k = L$, which yields the local PIR rate $1$. In contrast, the PIR capacity is $\Theta(1/\sqrt{N})$\cite{SGT23}. Intuitively, this is because, each server that the user interacts with has only one message, i.e., $|\mathcal{I}_n|=1$, $n \in [N-1]$.
\end{example}

\section{Main Results}
We start with some useful definitions. Then, we provide the capacity result on disjoint union of graphs, followed by results on important families of connected graphs. 
\begin{definition}[Disjoint union of graphs] 
    Graph $G=(V,E)$ is a disjoint union of $m$ graphs $G_i = (V_i,E_i)$, $i\in[m]$, if
    \begin{align}
        &V_i \cap V_j = \emptyset, \quad i,j\in [m],~ i\neq j,\\
        &V=\bigcup_{i=1}^m  V_i,  \quad E = \bigcup_{i=1}^m E_i.
    \end{align}
\end{definition}

\begin{definition}[Bipartite graphs]
    Graph $G=(V,E)$ is bipartite, if the vertex set $V$ is a disjoint union of two sets $V_1$ and $V_2$, and every edge $\{i,j\}\in E$, has $i\in V_1$ and $j\in V_2$. 
\end{definition}
Examples of bipartite graphs are cyclic graphs with even $N$, path graphs, star graphs and complete bipartite graphs. Given a graph $G=(V,E)$, its \emph{vertex cover} is a subset of vertices $V'\subseteq V$ such that every edge in $E$ has at least one end point in $V'$. Note that, $V_1$ and $V_2$ are vertex covers of $G$, which are also \emph{independent sets} of $G$, i.e., they share no common edge. 

\begin{definition}[Edge-transitive graphs]\label{def:edge_trans}
    Graph $G=(V,E)$ is edge-transitive, if for every pair of edges $e_1,e_2\in E$ there exists an automorphism that maps $e_1$ to $e_2$. 
\end{definition}
Examples of edge-transitive graphs are cyclic graphs, star graphs, complete graphs and complete bipartite graphs. 

\begin{theorem}\label{thm:disjoint_graphs_capacity}
    Let $G = \bigcup_{i=1}^m G_i$, where $G_i = (V_i, E_i)$. Let the capacity-achieving local PIR scheme on $G_i$ entail the message length $L_i$, and the expected download cost $\mathbb{E}[D_i]$  for the respective $|E_i|$ messages. Then, the local PIR capacity of $G$ is given by,
    \begin{align}\label{eq:disjoint_union_graphs}
        C(G)  = \frac{\sum_{i=1}^m  |E_i| \cdot L_i}{\sum_{i=1}^m |E_i| \cdot \mathbb{E}[D_i]}.
    \end{align}
\end{theorem}

\begin{Proof}
We present the proof for $m=2$, and the case of $m>2$ follows by induction. Let $E_1=[K_1]$ and $E_2=[K_1+1:K_1+K_2]$. Then, the scheme $\Pi$ can be described as follows. Given $k\in E_1\cup E_2$, the user follows the optimal local PIR scheme on $G_1$ if $k\in E_1$ and the optimal local PIR scheme on $G_2$ if $k\in E_2$, which yields the rate in \eqref{eq:disjoint_union_graphs}. Suppose, for the sake of contradiction, that there exists a scheme $\Pi'$ that achieves $R^{\Pi'}(G)> \frac{|E_1|L_1+|E_2|L_2}{|E_1|\mathbb{E}[D_1] + |E_2|\mathbb{E}[D_2]}$. Then, if $V_2=\emptyset$, we have $G=G_1$, and that $C(G)\geq R^{\Pi'}(G)>\frac{L_1}{\mathbb{E}[D_1]}$. This leads to a contradiction, since $C(G)=C(G_1)$ is at most $\frac{L_1}{\mathbb{E}[D_1]}$, hence, no such $\Pi'$ exists.
\end{Proof}

\begin{corollary}
    If the $m$ graphs $G_1,\ldots,G_m$ are identical, then $C(G) = C(G_1)$, i.e., there is no reduction in capacity. This gives multiplicative gain compared to the PIR capacity in the same setting, since $C(G)\geq C_{PIR}(G) = \frac{1}{m}C_{PIR}(G_1)$\cite{our_journal2025}.
\end{corollary}

In an edge-transitive graph $G$, for each $k\in [K]$, let $i$ and $j$ be the two servers where $W_k$ is replicated, i.e., $\{k\}= \mathcal{I}_i \cap \mathcal{I}_j$. Consider the subgraph $G_k=G(V_k,E_k) \subseteq G(V,E)$ with
\begin{align}
    V_k &=\{m\in [N]:  \mathcal{I}_m\cap(\mathcal{I}_i\cup \mathcal{I}_j)\neq \emptyset\}, \\
    E_k &= \{\ell \in [K]:\ell \in \mathcal{I}_i\cup \mathcal{I}_j\},
\end{align}
i.e., $V_k$ is the set of servers sharing at least one message with server $i$ or server $j$, including servers $i$ and $j$, and $E_k$ is the set of messages stored at server $i$ or server $j$. If $G$ is edge-transitive, $G_k$ is identical for all $k\in [K]$. 

\begin{theorem}\label{thm:edge_trans}
For edge-transitive graph $G$, and a fixed edge $k\in E$, where $W_k$ is replicated at servers $i$ and $j$, the local PIR capacity is lower bounded as 
\begin{align}\label{eq:thm 2_gen}
    C(G)\geq &\max_{{t_i \in [\deg(i)], t_j\in [\deg(j)]}}\bigg(\lambda(t_i,t_j)\Big(\frac{\deg(i)}{t_i}+t_i-1\Big)\notag \\
    &\quad+\left(1-\lambda(t_i,t_j)\right)\Big(\frac{\deg(j)}{t_j}+t_j-1\Big)\bigg)^{-1},
\end{align}
where
\begin{align}
    \lambda(t_i,t_j) = \frac{\binom{\deg(i)-1}{t_i-1}}{\binom{\deg(i)-1}{t_i-1}+\binom{\deg(j)-1}{t_j-1}}.
\end{align}
\end{theorem}
The proof of Theorem~\ref{thm:edge_trans} follows from the achievable scheme given in Section~\ref{sec:ach_edge-trans}.

\begin{remark}\label{rem:equal deg i and j}
    If $\deg(i)=\deg(j)=d$ for all $G_k$, by setting $t_i=t_j=t$, \eqref{eq:thm 2_gen} can be simplified to
    \begin{align}
           \! \! \! \!  C(G)\geq 
            & \max \left(\frac{1}{\frac{d}{\lfloor \sqrt{d}\rfloor}+\lfloor \sqrt{d}\rfloor-1},\frac{1}{\frac{d}{\lceil \sqrt{d}\rceil}+\lceil \sqrt{d}\rceil-1} \right).
    \end{align}
\end{remark}
The proof of Remark~\ref{rem:equal deg i and j} is presented in Appendix~\ref{proof_rmk2}.
\begin{corollary}
    From Theorem~\ref{thm:edge_trans}, we obtain the following rates:
    \begin{enumerate}
    \item For the cyclic graph $G=\mathbf{C}_N$, the subgraph $G_k=\mathbf{P}_4$, i.e., the path graph with $4$ vertices, where $\deg(i) = \deg(j) = 2$, which yields
        \begin{align}
            C(\mathbf{C}_N)\geq \frac{1}{2},
        \end{align}  
    which is equal to $C_{PIR}(\mathbf{C}_N)=\frac{2}{N+1}$ if $N=3$, and strictly greater for all $N \geq 4$.
    \item For the complete graph $G=\mathbf{K}_N$, $\deg(i) = \deg(j) = N-1$, which yields 
    \begin{align}
     C(\mathbf{K}_N)\geq \frac{1}{2\sqrt{N-1}},
    \end{align} 
    while $\left(\frac{4}{3}-o(1)\right)\frac{1}{N}\leq C_{PIR}(\mathbf{K}_N)\leq \frac{1}{N(e-2)}$, as recently established in \cite{gePIR}. 
    \item For the complete, balanced bipartite graph $G=\mathbf{K}_{\frac{N}{2},\frac{N}{2}}$ for even $N$, $\deg(i) = \deg(j) = \frac{N}{2}$, which yields
    \begin{align}\label{eq:bipartite_balanced}
    C(\mathbf{K}_{\frac{N}{2},\frac{N}{2}})\geq \frac{2}{\sqrt{N}},
    \end{align}
    while $\frac{4}{3N}\leq C_{PIR}(\mathbf{K}_{\frac{N}{2},\frac{N}{2}})\leq \frac{1}{N(e^{0.5}-1)}$, by the tightest lower and upper bounds, established in \cite{krishnan_graph} and \cite{gePIR}, respectively. 
    \end{enumerate}
\end{corollary}

\begin{theorem}\label{thm:cyclic_converse}
For the cyclic graph $\mathbf{C}_N$, we have
    \begin{align}
      C(\mathbf{C}_N)=\frac{1}{2}.  
    \end{align}
\end{theorem}
The achievability of Theorem~\ref{thm:cyclic_converse} follows from Theorem~\ref{thm:edge_trans} and the converse is derived in Section~\ref{sec:converse_cycle}. 

The next result focuses on bipartite graphs. The proof of Theorem~\ref{thm:ach_bipartite} is presented in Section~\ref{sec:ach_bipart}.

\begin{theorem}\label{thm:ach_bipartite}
    If $G=(V,E)$ is bipartite, with sets $V_1, V_2$, the local PIR capacity is lower bounded as
    \begin{align}
        C(G)\geq K\cdot \left(\min_{m\in \{1,2\}}\sum_{n\in V_m } \deg(n)^2\right)^{-1}.\label{eq:lbnd_bipartite}
    \end{align}
\end{theorem}

\begin{corollary}
From Theorem~\ref{thm:ach_bipartite}, we have the following rates:
\begin{enumerate}
    \item For the star graph $\mathbf{S}_N$, the scheme achieves
    \begin{align}
        C(\mathbf{S}_{N}) = 1,
    \end{align}
    by choosing $V_1=\{1,\ldots,N-1\}$ and $V_2=\{N\}$, since $\deg(n)=1$ for each $n\in V_1$, and $K=N-1$. This is the capacity since trivially, $C(G)\leq 1$ for any $G$.
    \item For the path graph $\mathbf{P}_N$, the scheme yields
    \begin{align}\label{eq:path_lbnd}
        C(\mathbf{P}_N)\geq \begin{cases}
            \frac{N-1}{2N-3}, & N \text{ even},\\
            \frac{N-1}{2N-4}, & N \text{ odd},
        \end{cases}
    \end{align}
    which is greater than $C_{PIR}(\mathbf{P}_N)=\frac{2}{N}$, by choosing $V_1$ as the odd, and $V_2$ as the even vertex indices in $[N]$.
\end{enumerate}
\end{corollary}

\begin{theorem}\label{thm:path_converse}
    For the path graph $\mathbf{P}_N$, we have 
    \begin{align}
        C(\mathbf{P}_N)\leq \frac{N-1}{2N-4}, \label{eq:path_ubnd}
    \end{align}
    which is tight for odd $N$ settling the capacity.
\end{theorem}
The proof of Theorem~\ref{thm:path_converse} relies on similar ideas as the converse proof of Theorem~\ref{thm:cyclic_converse}, and is provided in Appendix~\ref{proof:path_converse}. For odd $N$, Theorem~\ref{thm:path_converse} gives the capacity as the upper bound in \eqref{eq:path_ubnd} matches the rate in \eqref{eq:path_lbnd} for odd $N$.

\section{Achievability Proofs}
\subsection{{Proof of Theorem}~\ref{thm:edge_trans}}\label{sec:ach_edge-trans}
We present the scheme for a fixed $G_k$ where $\{k\}=\mathcal{I}_i\cap \mathcal{I}_j$ when $\theta = k$. By edge-transitivity of $G$, the same scheme applies to all $G_k$. The user communicates only with the servers in $V_k$ and downloads only the messages corresponding to $E_k$, i.e., $\mathcal{W}_i\cup \mathcal{W}_j$. We fix $t_i, t_j\in \mathbb{N}$, with $t_i \in [\deg(i)]$ and $t_j \in [\deg(j)]$. Let each message consist of 
\begin{align}\label{eq:subpacketization}
    L=\binom{\deg(i)-1}{t_i-1}+\binom{\deg(j)-1}{t_j -1} 
\end{align}
symbols. The user applies a private, uniformly random, independent permutation to the message symbols of each message in $\mathcal{W}_i\cup \mathcal{W}_j$, with $W_\ell(m)$ denoting the $m$th permuted symbol of $W_\ell$. To servers $i$ and $j$, the user sends queries for all possible $t_i$-sums of the messages in $\mathcal{W}_i$, and all possible $t_j$-sums of the messages in $\mathcal{W}_j$, respectively. In particular, let $\mathcal{A}_i$ denote the ordered set of all $t_i$-subsets of $\mathcal{I}_i$,  and $\mathcal{A}_j$ denote the ordered set of all $t_j$-subsets of $\mathcal{I}_j$, arranged in a fixed lexicographic order. Let $\mathcal{A}_i=\big(A^i_1,\ldots, A^i_{\binom{\deg(i)}{t_i}}\big)$ and $\mathcal{A}_j=\big(A^j_1,\ldots, A^j_{\binom{\deg(j)}{t_j}}\big)$. Then, for each $\ell \in \mathcal{I}_i$ and each $A_p^i\in \mathcal{A}_i$, such that $\ell\in A_p^i$, we define $\gamma_\ell(A_p^i)$ as
\begin{align}
    \gamma_\ell(A_p^i) = \sum_{x=1}^{p}{\mathds{1}\{\ell\in A^i_x\}},
\end{align}
i.e., the number of times that $\ell$ is contained in a subset, up to and including $A_p^i$. The same definition extends to $\mathcal{A}_j$. Then, for each $A\in \mathcal{A}_i$, the user downloads from server $i$, 
\begin{align}\label{msg_sum i}
    \sum_{\ell\in A} W_\ell(\gamma_\ell(A)).
\end{align}
From server $j$, for each  $A\in\mathcal{A}_j$, the user downloads 
\begin{align}\label{msg_sum j}
    \begin{cases}
        W_k\left(\binom{\deg(i)-1}{t_i-1}+\gamma_k(A)\right)+\sum_{\ell\in A\setminus\{k\}}W_\ell(\gamma_\ell(A)),\! \! \! & k\in A,\\
      \sum_{\ell\in A} W_\ell(\gamma_\ell(A)),& k\notin A.
    \end{cases} 
\end{align}
For each $\ell\in \mathcal{I}_i$, and each $\ell\in \mathcal{I}_j$, $\gamma_\ell(A)$ varies from 1 to $ \binom{\deg(i)-1}{t_i-1}$, and from 1 to $ \binom{\deg(j)-1}{t_j-1}$, respectively. From the servers, $V_k \setminus \{i,j\}$, the user downloads the interference message symbols, that appear in the summations \eqref{msg_sum i} and \eqref{msg_sum j}. Specifically, for $\ell\in \mathcal{I}_i\setminus\{k\}$, define $\mathcal{B}_{\ell,i} =\{B\in \mathcal{A}_i: \{\ell,k\}\subseteq B\}$, with $\mathcal{B}_{\ell,j}$ defined similarly with $\mathcal{I}_j$. Then, corresponding to $\ell\in \mathcal{I}_i\setminus \{k\}$ (or $\mathcal{I}_j\setminus\{k\}$), the user downloads the symbols, $\{W_\ell(\gamma_\ell(B)): B\in \mathcal{B}_{\ell,i} \text{ (or $  \mathcal{B}_{\ell,j}$)}\}$ from the server storing $W_\ell$. The total download is
\begin{align}
|&\mathcal{A}_i| + |\mathcal{A}_j| + \sum_{\ell\in \mathcal{I}_i\setminus \{k\}}|\mathcal{B}_{\ell,i}|+ \sum_{\ell\in \mathcal{I}_j\setminus \{k\}}|\mathcal{B}_{\ell,j}|\notag \\
&=\binom{\deg(i)}{t_i}+\binom{\deg(j)}{t_j}+(\deg(i)-1)\binom{\deg(i)-2}{t_i-2}\notag \\
&\quad+(\deg(j)-1)\binom{\deg(j)-2}{t_j-2}\\
&=\binom{\deg(i)-1}{t_i-1}\left(\frac{\deg(i)}{t_i}+t_i-1\right)\notag \\
&\quad+ \binom{\deg(j)-1}{t_j-1}\left(\frac{\deg(j)}{t_j}+t_j-1\right).\label{eq:download_cost} 
\end{align}
The ratio between \eqref{eq:subpacketization} and \eqref{eq:download_cost} gives the desired rate. The rate is maximized over $t_i$ and $t_j$ to obtain the largest lower bound. Privacy holds at servers $i$ and $j$ since the same scheme is followed for all $\theta$ in $\mathcal{I}_i$ and $\mathcal{I}_j$, respectively.

\subsection{Proof of Theorem \ref{thm:ach_bipartite}}\label{sec:ach_bipart}
Given $G$ bipartite, with sets $V_1$ and $V_2$ let 
\begin{align}
    m^* = \arg \min_{m\in \{1,2\}} \sum_{n\in V_m} \deg (n)^2.
\end{align}
For $\theta=k$, suppose $W_k$ is stored on servers $i$ and $j$. Assume without loss of generality that $i\in V_{m^*}$ and $j\notin V_{m^*}$. Then, the user queries for all messages from server $i$, downloading $\deg(i)$ messages and nothing at all from server $j$. The resulting total download is
\begin{align}
\sum_{k=1}^K D_k &= 
\sum_{n \in V_{m^*}}\deg(n)L \left(\sum_{k=1}^K \mathds{1}\{k\in \mathcal{I}_n\}\right)\\
&= \sum_{n\in V_{m^*}} \deg(n)^2 L.\label{eq:d_total_bipartite}
\end{align}
The ratio between $KL$ and \eqref{eq:d_total_bipartite} gives the desired rate.

\section{Converse Proof for Cyclic Graphs}\label{sec:converse_cycle}
We start with the following lemmas, which any feasible local PIR scheme should satisfy. The first lemma is a consequence of local user privacy, while the second lemma follows from the fact that $A_n^{[k]}$ is independent of $\mathcal{Q}$ given $Q_n^{[k]}$. The proofs are given in Appendices~\ref{proof:lem1} and \ref{proof:lem2}, respectively.

\begin{lemma}\label{lem:user_priv}
    For any $\mathcal{J}\subseteq [K]$, define the message set $W_{\mathcal{J}} = \{W_{\ell}:\ell\in \mathcal{J}\}$. Then, we have
    \begin{align}
        I(\theta; A_n^{[\theta]}|Q_n^{[\theta]}, W_{\mathcal{J}}) = 0, \quad \theta \in \mathcal{I}_n, ~n\in [N].
    \end{align}
\end{lemma}
\begin{lemma}\label{lem:answer_indep_randomness_given_query}
    For any subset $\mathcal{J}\subseteq [K]$, any $k\in [K]$, $n\in [N]$
    \begin{align}
        H(A_n^{[k]}|W_{\mathcal{J}}, Q_n^{[k]})&=H(A_n^{[k]}|W_{\mathcal{J}},Q_n^{[k]},\mathcal{Q})\label{eq:a}\\
        &=H(A_n^{[k]}|W_{\mathcal{J}},\mathcal{Q})\label{eq:b}.
    \end{align}
\end{lemma}

We now prove Theorem~\ref{thm:cyclic_converse}. Let $\mathcal{W}_n = \{W_n, W_{n-1}\}$, where the subtraction is taken modulo-$N$. Moreover, throughout the proof, we assume that the additions and subtractions are performed modulo-$N$, i.e., $W_0 = W_N$. Then, for $\theta = k$, the download cost $D_k$ can be bounded as
\begin{align}
    D_k &\geq \sum_{n=1}^N H(A_n^{[k]}|\mathcal{Q})\label{eq:download_first step}\geq H(A_{[N]}^{[k]}|\mathcal{Q})\\
    &=   H(W_k,A_{[N]}^{[k]}|\mathcal{Q})\label{eq:follows from decodability}\\   
    &= H(W_k|\mathcal{Q})+  H(A_{[N]}^{[k]}|W_k,\mathcal{Q})\\
    & = L+ I(\mathcal{W}\setminus \{W_k\};A_{[N]}^{[k]}|W_k,\mathcal{Q}),\label{eq:all answers deterministic}
\end{align}
where \eqref{eq:follows from decodability} follows since $H(W_k|A_{[N]}^{[k]},\mathcal{Q})=0$, and \eqref{eq:all answers deterministic} holds since $H(A_{[N]}^{[k]}|\mathcal{W},\mathcal{Q})=0$. Next, we lower bound the second term in \eqref{eq:all answers deterministic}, which is the interference in the answers,
\begin{align}
    &I(\mathcal{W}\setminus \{W_k\};A_{[N]}^{[k]}|W_k,\mathcal{Q})\notag \\
    &\geq I(W_{k-1}, W_{k+1} ; A_{[N]}^{[k]}|W_k, \mathcal{Q}) \\
    &= I(W_{k-1};A_{[N]}^{[k]}|W_k,\mathcal{Q}) d + I(W_{k+1};A_{[N]}^{[k]}|W_k, W_{k-1},\mathcal{Q})\\
    &\geq I(W_{k-1};A_{k}^{[k]}|W_k,\mathcal{Q})+ I(W_{k+1};A_{k+1}^{[k]}|W_k, W_{k-1},\mathcal{Q})\\
    &= I(W_{k-1};A_{k}^{[k-1]}|W_k,\mathcal{Q})+ I(W_{k+1};A_{k+1}^{[k+1]}|W_k, W_{k-1},\mathcal{Q})\label{eq:change_index_cycle}\\
    &=H(A_k^{[k-1]}|W_k,\mathcal{Q})+ H(A_{k+1}^{[k+1]}|W_k,  W_{k-1},\mathcal{Q}),\label{eq:cycle_answer_deterministic}
\end{align}
where \eqref{eq:change_index_cycle} follows from Lemma~\ref{lem:user_priv} and Lemma~\ref{lem:answer_indep_randomness_given_query}, and \eqref{eq:cycle_answer_deterministic} follows since $H(A_n^{[k]}|W_n, W_{n-1},\mathcal{Q}) = 0$ for all $n, ~k\in [N]$. Summing $D_k-L$ over all $k\in [N]$, we have,
\begin{align}
    &\sum_{k=1}^N  D_k - NL\notag \\
    &\geq \sum_{k=1}^N H(A_k^{[k-1]}|W_k,\mathcal{Q}) + H(A_{k+1}^{[k+1]}|W_k,  W_{k-1},\mathcal{Q})\\
    &= \sum_{k=1}^N H(A_k^{[k]}|W_{k-1},W_{k-2},\mathcal{Q})+ H(A_{k+1}^{[k]}|W_{k+1},\mathcal{Q})\label{eq:rearranging terms with same superscript}\\
    &\geq \sum_{k=1}^N H(A_k^{[k]}, A_{k+1}^{[k]}|W_{k-2},W_{k-1}, W_{k+1},\mathcal{Q})\\
    &\geq \sum_{k=1}^N H(A_k^{[k]}, A_{k+1}^{[k]}|\mathcal{W}\setminus \{W_k\},\mathcal{Q})\\
    &=\sum_{k=1}^N H(W_k,A_k^{[k]}, A_{k+1}^{[k]}|\mathcal{W}\setminus \{W_k\},\mathcal{Q})\label{eq:decodability of Wk}\\
    &= \sum_{k=1}^N H(W_k|\mathcal{W}\setminus \{W_k\},\mathcal{Q}) + H(A_k^{[k]}, A_{k+1}^{[k]}|\mathcal{W},\mathcal{Q})\label{eq:answers_deterministic}\\
    &= NL,
\end{align}
where \eqref{eq:rearranging terms with same superscript} follows by collecting the answers with the same index $k$, \eqref{eq:decodability of Wk} holds from decodability of $W_k$ since the answers $A^{[k]}_{[N]\setminus \{k,k+1\}}$ are deterministic functions of $\mathcal{W}\setminus \{W_k\}$ and $\mathcal{Q}$. Finally, the second term of \eqref{eq:answers_deterministic} is zero, completing the proof.

\newpage

\bibliographystyle{unsrt}
\bibliography{references.bib}

@ARTICLE{SGT23,
  author={Sadeh, B. and Gu, Y. and Tamo, I.},
  journal={IEEE Trans. Inf. Forensics Security}, 
  title={Bounds on the Capacity of Private Information Retrieval Over Graphs}, 
  year={2023},
  month={November},
  volume={18},
  number={},
  pages={261--273},
  }

@INPROCEEDINGS{BU19,
  author={Banawan, K. and Ulukus, S.},
  booktitle={IEEE ISIT}, 
  title={Private Information Retrieval from Non-Replicated Databases}, 
  year={2019},
  month={July},
  volume={},
  number={},
  doi={10.1109/ISIT.2019.8849399}
  }

@ARTICLE{SJ17,
  author={Sun, H. and Jafar, S. A.},
  journal={IEEE Trans. Inf. Theory}, 
  title={The Capacity of Private Information Retrieval}, 
  year={2017},
  month={March},
  volume={63},
  number={7},
  pages={4075--4088},
  doi={10.1109/TIT.2017.2689028}
}

@inproceedings{YJ23,
    author = {Yao, Y. and Jafar, S. A.},
    title = {The Capacity of 4-Star-Graph {PIR}},
    booktitle = {IEEE ISIT},
    year = {2023},
    month = {June}
}

@article{c_spir,
    title={The capacity of symmetric private information retrieval},
    author={H. Sun and S. A. Jafar},
    journal={IEEE Trans. Inf. Theory},
    volume={65},
    number={1},
    pages={322--329},
    year={2018},
    month={June}}

@article{C_SETPIR,
    title={The capacity of private information retrieval with eavesdroppers},
    author={Q. Wang and H. Sun and M. Skoglund},
    journal={IEEE Trans. Inf. Theory},
    volume={65},
    number={5},
    pages={3198--3214},
    year={2018},
    month={December}}

@article{csa,
    title={Cross subspace alignment and the asymptotic capacity of {$X$}-secure {$T$}-private information retrieval},
    author={Z. Jia and H. Sun and S. A. Jafar},
    journal={IEEE Trans. Inf. Theory},
    volume={65},
    number={9},
    pages={5783--5798},
    year={2019},
    month = {May},
    }

@article{coded_colluding_2017,
   title={Private Information Retrieval from Coded Databases with Colluding Servers},
   author={Freij-Hollanti, R. and Gnilke, O. W. and Hollanti, C. and Karpuk, D. A.},
   journal={SIAM J. Appl. Algebra and Geometry},
   volume={1},
   number={1},
   pages={647-664},
   year={2017},
   month={}}

@article{chor,
    title={Private information retrieval},
    author={B. Chor and E. Kushilevitz and O. Goldreich and M. Sudan},
    journal={Journal of the ACM},
    volume={45},
    number={6},
    pages={965-981},
    year={1998},
    month={November}}

@article{arbitrarycollusion,
    title={The capacity of private information retrieval under arbitrary collusion patterns for replicated databases},
    author={X. Yao and N. Liu and W. Kang},
    journal={IEEE Trans. Inf. Theory},
    volume={67},
    number={10},
    pages={6841-6855},
    year={2021},
    month={July}}

@article{graphbased_pir,
    title={Private information retrieval in graph-based replication systems},
    author={N. Raviv and I. Tamo and E. Yaakobi},
    journal={IEEE Trans. Inf. Theory},
    volume={66},
    number={6},
    pages={3590-3602},
    year={2019},
    month={November}}

@article{byzantine_tpir,
    title={The capacity of private information retrieval from {B}yzantine and colluding databases},
    author={K. Banawan and S. Ulukus},
    journal={IEEE Trans. Inf. Theory},
    volume={65},
    number={2},
    pages={1206-1219},
    year={2018},
    month={September}}

@article{nan_eaves,
    title={The capacity of symmetric private information retrieval under arbitrary collusion and eavesdropping patterns},
    author={J. Cheng and N. Liu and W. Kang and Y. Li},
    journal={IEEE Trans. Inf. Forensics Security},
    volume={17},
    pages={3037-3050},
    year={2022},
    month={August}}

@article{sun_eaves,
    title={The capacity of private information retrieval with eavesdroppers},
    author={Q. Wang and H. Sun and M. Skoglund},
    journal={IEEE Trans. Inf. Theory},
    volume={65},
    number={5},
    pages={3198-3214},
    year={2018},
    month={December}}

@article{colluding,
    title ={The Capacity of Robust Private Information Retrieval With Colluding Databases},
    author={H. Sun and S. A. Jafar},
    journal={IEEE Trans. Inf. Theory},
    year={2018},
    month={April},
    volume={64},
    number={4},
    pages={2361-2370}}

@ARTICLE{skoglund_mds_spir,
  author={Wang, Q. and Skoglund, M.},
  journal={IEEE Trans. Inf. Theory}, 
  title={Symmetric Private Information Retrieval from {MDS} Coded Distributed Storage With Non-Colluding and Colluding Servers}, 
  year={2019},
  volume={65},
  number={8},
  pages={5160-5175},
  month = {March}}

@ARTICLE{zhusheng_spir_pir,
  author={Wang, Z. and Ulukus, S.},
  journal={IEEE J. Sel. Areas Inf. Theory}, 
  title={Symmetric Private Information Retrieval at the Private Information Retrieval Rate}, 
  year={2022},
  month={October},
  volume={3},
  number={2},
  pages={350-361},
  doi={10.1109/JSAIT.2022.3188610}}

@inproceedings{meel_multi_pir,
  title={Private Information Retrieval on Multigraph-Based Replicated Storage},
  author={Meel, S. and Kong, X. and Maranzatto, T. J. and Tamo, I. and Ulukus, S.}, 
  booktitle = {IEEE ISIT},
  year = {2025},
  month = {June}
}

@inproceedings{nomeir_asymp_bspir,
  title={The Asymptotic Capacity of {B}yzantine Symmetric Private Information Retrieval and Its Consequences},
  author={Nomeir, M. and Aytekin, A. and Ulukus, S.},
  booktitle = {IEEE ISIT},
  year = {2025},
  month = {June}
}

@ARTICLE{pir_spir_adversaries,
  author={Wang, Q. and Skoglund, M.},
  journal={IEEE Trans. Inf. Theory}, 
  title={On {PIR} and Symmetric {PIR} From Colluding Databases With Adversaries and Eavesdroppers}, 
  year={2019},
  volume={65},
  number={5},
  pages={3183-3197},
  month={October}}

@article{our_journal2025,
      title={New Capacity Bounds for {PIR} on Graph and Multigraph-Based Replicated Storage}, 
      author={X. Kong and S. Meel and T. J. Maranzatto and I. Tamo and S. Ulukus},
      journal = {IEEE Trans. Inf. Theory},
      year={2026},
      month = {January},
      volume ={72},
      number = {1},
      pages={691-709},
}

@misc{gePIR,
  title={Private Information Retrieval over Graphs},
  author={Ge, G. and Wang, H. and Xu, Z. and Zhang, Y.},
  note={Available online at arXiv:2509.26512},
  year={2025}
}

@inproceedings{nomeirasymmetric,
    title={Asymmetric ${X}$-Secure ${T}$-Private Information Retrieval: More Databases is Not Always Better}, 
    author={M. Nomeir and S. Vithana and S. Ulukus},
    booktitle={CISS},  
    year={2024},
    month= {March}}

@misc{krishnan_graph,
  title={Private Information Retrieval for Graph-based Replication with Minimal Subpacketization},
  author={Shanbhag, V. and Krishnan, P.},
  note={Available online at arXiv:2601.09957},
  year={2026}
}

@inproceedings{semantic_tpir,
    author = {M. Nomeir and A. Aytekin and S. Ulukus},
    title = {The Capacity of Semantic Private Information Retrieval with Colluding Servers},
    booktitle = {IEEE GLOBECOM},
    year = {2025},
    month = {December}
}

\clearpage

\appendix
\subsection{Proof of Remark~\ref{rem:equal deg i and j}}\label{proof_rmk2}
The expression that we minimize is
\begin{align}\label{eq:convex_comb}
    \lambda(t_i,t_j)\left(\frac{d}{t_i}+t_i-1\right)+(1-\lambda(t_i,t_j))\left(\frac{d}{t_j}+t_j -1\right),
\end{align}
where 
\begin{align}
\lambda(t_i,t_j) & = \frac{\binom{d-1}{t_i-1}}{\binom{d-1}{t_i-1}+\binom{d-1}{t_j-1}}\in (0,1).
\end{align}
Assume without loss of generality, that $\frac{d}{t_i}+t_i-1\leq \frac{d}{t_j}+t_j -1$. Now, for any $\lambda(t_i,t_j)\in (0,1)$, 
\begin{align}
    \lambda(&t_i,t_j)\left(\frac{d}{t_i}+t_i-1\right)+(1-\lambda(t_i,t_j))\left(\frac{d}{t_j}+t_j -1\right)\notag \\
    &\geq \min\left(\frac{d}{t_i}+t_i-1, \frac{d}{t_j}+t_j-1\right)\\
    &= \frac{d}{t_i}+t_i-1,\label{eq:denominator}
\end{align}
The minimum value of \eqref{eq:denominator} is achieved by setting $t_i = t_j =t $ in \eqref{eq:convex_comb}, and solving 
\begin{align}\label{solve_opt}
      \max_{t\in [d]} \frac{1}{\frac{d}{t}+t-1}.  
\end{align}
The solution of \eqref{solve_opt} yields the optimal value of $t\in \Big\{\frac{d}{\lceil\sqrt{d}\rceil}, \frac{d}{\lfloor\sqrt{d}\rfloor}\Big\}$, which completes the proof.

\subsection{Proof of Lemma~\ref{lem:user_priv}}\label{proof:lem1}
It is equivalent to showing that for any $k,k'\in \mathcal{I}_n$, $k\neq k'$,
\begin{align}
    H(A_n^{[k]}|Q_n^{[k]},W_\mathcal{J})=H(A_n^{[k']}|Q_n^{[k']},W_\mathcal{J}).
\end{align}
Let $\mathcal{W}_n' = \mathcal{W}_n \cap W_{\mathcal{J}}$. Note that,
\begin{align}
    &H(A_n^{[k]}|Q_n^{[k]},W_\mathcal{J})\notag\\
    &=   H(A_n^{[k]}|Q_n^{[k]},\mathcal{W}_n')-I(W_{\mathcal{J}}\setminus \mathcal{W}_n';A_n^{[k]}|Q_n^{[k]},\mathcal{W}_n')\\
    &=H(A_n^{[k]}|Q_n^{[k]},\mathcal{W}_n'),\label{eq:ans indep of non intersecting msgs}
\end{align}
since
\begin{align}
    I&(W_{\mathcal{J}}\setminus \mathcal{W}_n';A_n^{[k]}|Q_n^{[k]},\mathcal{W}_n')\notag\\
    &\leq  I(W_{\mathcal{J}}\setminus \mathcal{W}_n';A_n^{[k]}, \mathcal{W}_n\setminus \mathcal{W}_n'|Q_n^{[k]},\mathcal{W}_n')\\
    &= I(W_{\mathcal{J}}\setminus \mathcal{W}_n';\mathcal{W}_n\setminus \mathcal{W}_n'|Q_n^{[k]},\mathcal{W}_n')\notag\\
    &\quad+I(W_{\mathcal{J}}\setminus \mathcal{W}_n';A_n^{[k]}|Q_n^{[k]},\mathcal{W}_n)\\
    &\leq I(W_{\mathcal{J}}\setminus \mathcal{W}_n';\mathcal{W}_n|Q_n^{[k]})+ H(A_n^{[k]}|Q_n^{[k]},\mathcal{W}_n)\\
    &=0.
\end{align}
From \eqref{eq:user privacy new} and \eqref{answer_function}, for every $n$ and $\theta\in \mathcal{I}_n$,
\begin{align}
    0=I(\theta;Q_n^{[\theta]},A_n^{[\theta]},\mathcal{W}_n) \geq I(\theta;Q_n^{[\theta]},A_n^{[\theta]},\mathcal{W}_n').\label{eq:conseq of user priv}
\end{align}
Finally, combining \eqref{eq:ans indep of non intersecting msgs} and \eqref{eq:conseq of user priv} proves the lemma.

\subsection{Proof of Lemma~\ref{lem:answer_indep_randomness_given_query}}\label{proof:lem2}
For \eqref{eq:a}, we show that $I(A_n^{[k]};\mathcal{Q}|W_{\mathcal{J}}, Q_n^{[k]})=0$. Denoting $\mathcal{W}_n'$ as $\mathcal{W}_n \cap W_{\mathcal{J}}$, we have
\begin{align}
    I(&A_n^{[k]};\mathcal{Q}|W_{\mathcal{J}}, Q_n^{[k]}) \notag\\
    \leq& I(A_n^{[k]},\mathcal{W}_{n}\setminus\mathcal{W}_n';\mathcal{Q}|W_{\mathcal{J}}, Q_n^{[k]})\\
    =& I(\mathcal{W}_{n}\setminus\mathcal{W}_n';\mathcal{Q}|W_{\mathcal{J}}, Q_n^{[k]})\!+\!I(A_n^{[k]};\mathcal{Q}|\mathcal{W}_n, W_{\mathcal{J}}\setminus \mathcal{W}_n',Q_n^{[k]})\label{eq:last term is zero}\\
    \leq&I(\mathcal{W}_{n}\setminus\mathcal{W}_n';\mathcal{Q}|W_{\mathcal{J}}, Q_n^{[k]}) + I(W_{\mathcal{J}};\mathcal{Q}|Q_n^{[k]})\label{eq:follows by answer generation}\\
    =&I(\mathcal{W}_n, W_{\mathcal{J}}\setminus \mathcal{W}_n';\mathcal{Q}|Q_n^{[k]})\\
    \leq& I(\mathcal{W}_n, W_{\mathcal{J}}\setminus \mathcal{W}_n';\mathcal{Q},Q_n^{[k]})=0,
\end{align}
where \eqref{eq:follows by answer generation} follows since the second term of \eqref{eq:last term is zero} is zero because $H(A_n^{[k]}|\mathcal{W}_n,W_{\mathcal{J}}\setminus \mathcal{W}_n',Q_n^{[k]})=0$, and the last equality follows by \eqref{query-msg_indep}. The equality \eqref{eq:b} holds since $H(Q_n^{[k]}|\mathcal{Q}) = 0$.

\subsection{Proof of Theorem~\ref{thm:path_converse}}\label{proof:path_converse}
Given $\mathbf{P}_N$, let the storage of the servers be 
\begin{align}
    \mathcal{W}_n =
    \begin{cases}
         \{W_1\}, & n=1,\\
         \{W_n, W_{n-1}\}, & n\in [2:N-1],\\
         \{W_{N-1}\}, &n=N.
    \end{cases}
\end{align} 
For $\theta = k$, following steps as in \eqref{eq:download_first step}--\eqref{eq:all answers deterministic}, we obtain
\begin{align}
    D_k &\geq L+ I(\mathcal{W}\setminus \{W_k\};A_{[N]}^{[k]}|W_k,\mathcal{Q}).
\end{align}
Next, we lower bound the second term. First, for $k=1$,
\begin{align}
   I(\mathcal{W}\setminus \{W_1\};A_{[N]}^{[1]}|W_1,\mathcal{Q})&\geq I(W_{2} ; A_{[N]}^{[1]}|W_1, \mathcal{Q})\\
   &\geq I(W_{2} ; A_2^{[1]}|W_1, \mathcal{Q})\\
   &=H(A_2^{[1]}|W_1, \mathcal{Q})\\
   &= H(A_2^{[2]}|W_1, \mathcal{Q}).
\end{align}
Similarly, for $k=N-1$, we have
\begin{align}
   I(\mathcal{W}\setminus \{W_{N-1}\};A_{[N]}^{[N-1]}|W_{N-1},\mathcal{Q})&\geq 
   H(A_{N-1}^{[N-2]}|W_{N-1}, \mathcal{Q}).
\end{align}
For $k\in [2:N-2]$, we have
\begin{align}
    &I(\mathcal{W}\setminus \{W_k\};A_{[N]}^{[k]}|W_k,\mathcal{Q})\notag\\
    &\geq I(W_{k-1}, W_{k+1} ; A_{[N]}^{[k]}|W_k, \mathcal{Q})\\
    & = I(W_{k-1};A_{[N]}^{[k]}|W_k,\mathcal{Q})+ I(W_{k+1};A_{[N]}^{[k]}|W_k, W_{k-1},\mathcal{Q})\\
    &\geq I(W_{k-1};A_{k}^{[k]}|W_k,\mathcal{Q})+ I(W_{k+1};A_{k+1}^{[k]}|W_k, W_{k-1},\mathcal{Q})\\
    &=  I(W_{k-1};A_{k}^{[k-1]}|W_k,\mathcal{Q})+ I(W_{k+1};A_{k+1}^{[k+1]}|W_k, W_{k-1},\mathcal{Q})\label{eq:path_follows from privacy}\\
    &=H(A_k^{[k-1]}|W_k,\mathcal{Q}) + H(A_{k+1}^{[k+1]}|W_k,  W_{k-1},\mathcal{Q}).
\end{align}
where \eqref{eq:path_follows from privacy} is due to Lemmas~\ref{lem:user_priv} and \ref{lem:answer_indep_randomness_given_query}. Now, summing $D_k-L$ over all $k\in [N-1]$, we have
\begin{align}
    \sum_{k=1}^{N-1}&D_k - (N-1)L\notag \\
    \geq&  H(A_2^{[2]}|W_1,\mathcal{Q})\notag\\
    &+\sum_{k=2}^{N-2} H(A_k^{[k-1]}|W_k,\mathcal{Q}) + H(A_{k+1}^{[k+1]}|W_k,  W_{k-1},\mathcal{Q}) \notag\\
    &+H(A_{N-1}^{[N-2]}|W_{N-1},\mathcal{Q})\\
    \geq & H(A_2^{[2]}|W_1,\mathcal{Q})+H(A_3^{[2]}|W_3,\mathcal{Q})\notag\\
    &+ \sum_{k=3}^{N-2} H(A_k^{[k]}|W_{k-1},W_{k-2},\mathcal{Q})+H(A_{k+1}^{[k]}|W_{k+1},\mathcal{Q})\label{eq:path_dropped and rearranged}\\
     \geq& H(A_2^{[2]},A_3^{[2]}|W_1,W_3,\mathcal{Q})\notag\\
    &+\sum_{k=3}^{N-2} H(A_k^{[k]}, A_{k+1}^{[k]}|W_{k-2}, W_{k-1}, W_{k+1},\mathcal{Q})\\
    \geq&  H(A_2^{[2]},A_3^{[2]}|\mathcal{W}\setminus \{W_2\},\mathcal{Q})\notag\\
    &+\sum_{k=3}^{N-2} H(A_k^{[k]}, A_{k+1}^{[k]}|\mathcal{W}\setminus \{W_k\},\mathcal{Q})\\
    =&H(W_2,A_2^{[2]},A_3^{[2]}|\mathcal{W}\setminus \{W_2\},\mathcal{Q})\notag\\
    &+\sum_{k=3}^{N-2} H(W_k,A_k^{[k]}, A_{k+1}^{[k]}|\mathcal{W}\setminus \{W_k\},\mathcal{Q})\\
    =& H(W_2|\mathcal{W}\setminus \{W_2\},\mathcal{Q})  +H(A_2^{[2]},A_3^{[2]}|\mathcal{W},\mathcal{Q})\notag \\
    &+\sum_{k=3}^{N-2} H(W_k|\mathcal{W}\setminus \{W_k\},\mathcal{Q})+ H(A_k^{[k]}, A_{k+1}^{[k]}|\mathcal{W},\mathcal{Q})\label{eq:path_holds by decodability}\\
    =& (N-3)L.
\end{align}
where \eqref{eq:path_dropped and rearranged} follows by dropping the terms $H(A_2^{[1]}|W_2,\mathcal{Q})$ and $H(A_{N-1}^{[N-1]}|W_{N-3}, W_{N-2})$, and re-arranging the remaining terms, and \eqref{eq:path_holds by decodability} holds by decodability, completing the proof.
\end{document}